\documentclass{article}
\usepackage{spconf,amsmath,graphicx}
\usepackage{multirow}
\usepackage{footnote}
\usepackage{array}
\usepackage{cite}

\title{ENHANCING  MULTILINGUAL SPEECH RECOGNITION THROUGH LANGUAGE PROMPT TUNING AND FRAME-LEVEL LANGUAGE ADAPTER}
\name{Song Li, Yongbin You, Xuezhi  Wang, Ke Ding, Guanglu Wan}
\address{Meituan\\
 \small{\texttt{ \{lisong39, dingke02, wanguanglu\}@meituan.com}}
  }

\begin{document}
%
\maketitle
\begin{abstract}
Multilingual intelligent assistants, such as ChatGPT, have recently gained popularity. To further expand the applications of multilingual artificial intelligence (AI) assistants and facilitate international communication, it is essential to enhance the performance of multilingual speech recognition, which is a crucial component of speech interaction. In this paper, we propose two simple and parameter-efficient methods: $language$ $prompt$  $tuning$ and $frame$-$level $ $language$ $ adapter$, to  respectively enhance language-configurable and language-agnostic multilingual speech  recognition. Additionally, we explore the feasibility of integrating these two approaches using parameter-efficient fine-tuning methods. Our experiments demonstrate significant performance improvements across seven languages using our proposed methods.
\end{abstract}

\section{Introduction}
\label{sec:intro}
Recently, multilingual AI assistants represented by ChatGPT have brought great convenience to people's work and life. As an important component of speech interaction, multilingual speech recognition has also received much attention from researchers to promote wider applications of speech interaction\cite{pratap2020massively,watanabe2017language,toshniwal2018multilingual,kannan2019large,kwon2023mole,punjabi2020streaming,zhang2022streaming}. The multilingual speech recognition systems mentioned in this paper refer to architectures that can recognize multiple languages using a single model. However, this all-in-one architecture may decrease speech recognition performance for certain languages due to language confusion. Generally, low-resource languages  benefit from the architecture, while high-resource languages may experience a decline\cite{pratap2020massively}. Therefore, how to improve the language discrimination ability  to enhance the performance of multilingual speech recognition systems has become a key research topic.

Currently, there are two mainstream approaches to enhancing the language discrimination ability of  multilingual speech recognition: incorporating language information and self-supervised training. For the former, language information can be directly integrated into the model or additional components can be introduced to provide language bias and improve the performance of specific languages. Ref.\cite{punjabi2020streaming,zhang2022streaming} introduced an additional language identification (LID) module to predict language information, while Ref.\cite{watanabe2017language} treated language information as a special textual token and concatenated it to the input of the decoder of the autoregressive speech recognition model, achieving joint modeling of speech recognition and language identification. Ref.\cite{toshniwal2018multilingual} provided language information directly as prior information to speech recognition models, this can be achieved by encoding language information as a one-hot vector or embedding and concatenating it with acoustic features. Alternatively, language-specific sub-modules can be built based on prior language information, such as language-specific adapters\cite{kannan2019large}, mixture of experts\cite{kwon2023mole}, and decoders\cite{pratap2020massively}. For the latter, self-supervised learning-based methods employ a significant corpus of multilingual audio only data for pre-training. By utilizing contrastive learning or masking language model loss function, the model acquires language discrimination ability and rich multilingual acoustic semantic knowledge, which is subsequently fine-tuned with multilingual annotated data\cite{conneau2020unsupervised}.

Language-agnostic and language-configurable speech recognition are two major application scenarios of multilingual speech recognition systems. The former is suitable for scenarios where the language is unknown in advance, while the latter provides users with a switch to select the language to be recognized. In previous research, these two scenarios were usually studied separately. In this paper, we attempt to explore simple and parameter-efficient methods to study both scenarios simultaneously and investigate the feasibility of integrating them. Specifically, for language-configurable scenarios, we propose simple and efficient language prompt tuning based methods to achieve language adaptation. For language-agnostic scenarios, we propose using a frame-level language adapter to enhance the model's ability to distinguish between languages.   Finally, we  explore the possibility of merging the models for both scenarios into a  unified model using parameter-efficient fine-tuning methods.

The rest of the paper is organized as follows. In Section 2, the details of our proposed  parameter-efficient methods  are described. In Section 3, we introduce specific experimental details. Experimental results are presented in Section 4. Finally, the paper is concluded in Section 5. 

\section{Proposed technology}
\vspace{-0.1cm}
\label{sec:format}
In this paper, we focus on exploring methods to enhance the performance of CTC-based multilingual speech recognition through the utilization of language information.  Inspired by large language models (LLM), we first propose prompt tuning to improve language-configurable scenarios. Next, we introduce the frame-level language adapter to enhance language-agnostic scenarios. Finally, we merge these strategies into a unified model using parameter-efficient fine-tuning methods.

\vspace{-0.2cm}
\subsection{Language Prompt Tuning }
\label{ssec:subhead}
\vspace{-0.1cm}
A common approach to leverage language information is to encode it as a one-hot vector or embedding\cite{toshniwal2018multilingual} , and then concatenate it with acoustic features to form the input for a multilingual speech recognition model. This can be achieved using the $Concat$ method described in Figure 1(a), or the $Add$ method described in Figure 1(b). For the $Add$ method, as shown in Figure 1(c), an attention mechanism  can also be used to obtain the weights of acoustic features and language embedding, followed by a weighted sum. 

\vspace{-0.4cm}
\begin{figure}[h]
  \centering
  \includegraphics[width=0.8\linewidth]{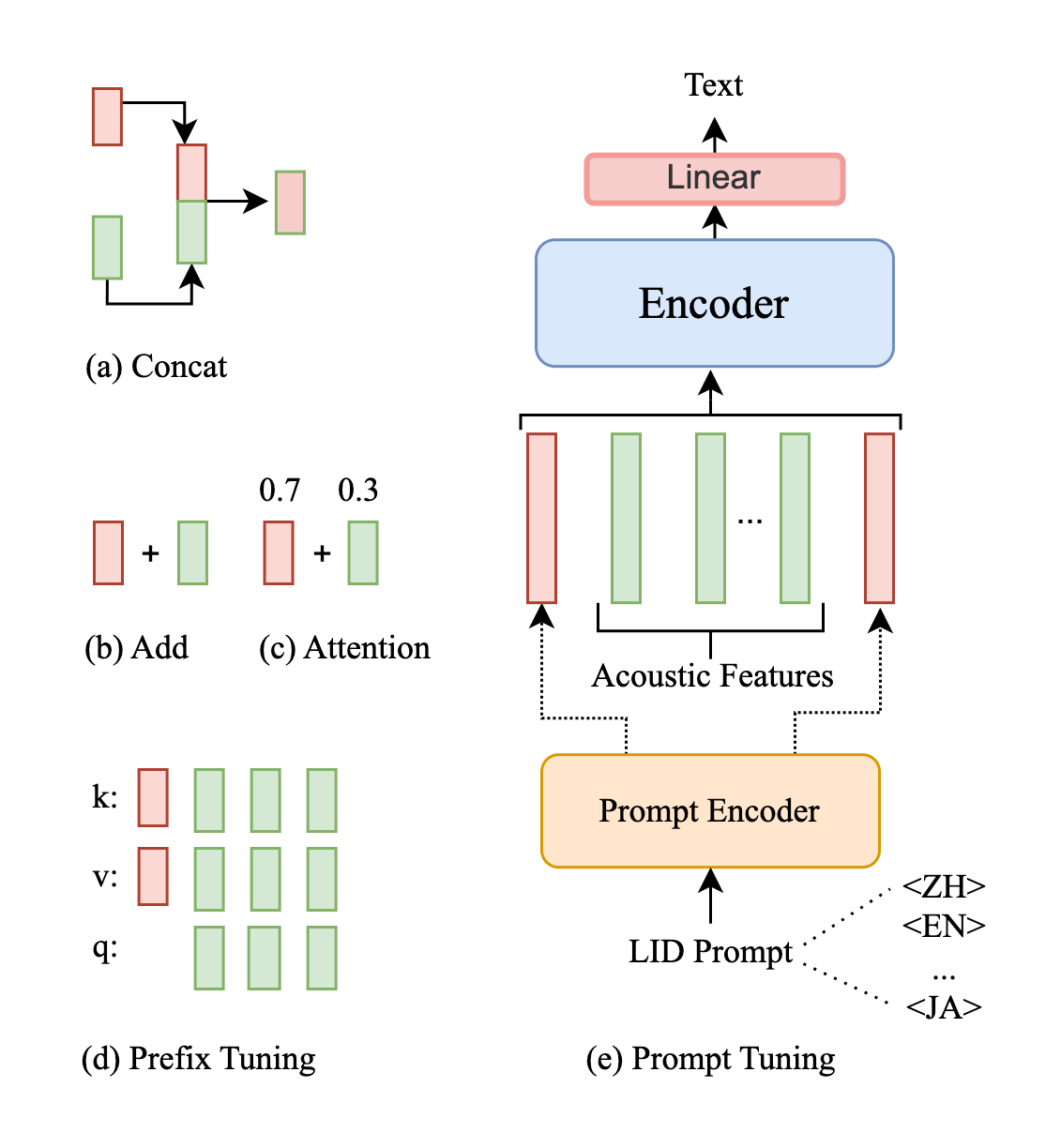}
  \vspace{-0.5cm}
  \caption{Methods of utilizing language information.}
  \label{fig:speech_production}
\end{figure}

Recently, multilingual intelligent assistants such as ChatGPT have introduced new ideas for developing multi-task unified models. By providing different prompts to the model, a single model can perform various tasks. Inspired by this, we attempt to introduce prompt tuning\cite{lester2021power} into CTC-based end-to end multilingual speech recognition. As described in Figure 1(e), we use a prompt encoder to encode language information (in the form of discrete LIDs) into a prompt embedding with the same dimension as the acoustic features. The prompt embedding serves as a language prompt, guiding the model to accurately recognize specific languages. We can position the prompt embedding as a prefix or suffix to the acoustic features using the transformer encoder as the acoustic encoder.  Another implementation of prompt tuning is prefix tuning\cite{li2021prefix} , as shown in Figure 1(d), which concatenates the prompt embedding to the key and value of each transformer layer. Compared to the $Concat$ and $Add$ methods, prompt tuning enables adaptive allocation of language information to each frame at every layer of the transformer encoder through self-attention mechanism, providing greater flexibility for the model to utilize language information at each layer.

\vspace{-0.1cm}
\subsection{Frame-level Language Adapter}
\label{ssec:subhead}
\vspace{-0.1cm}
To enhance a model's language discrimination ability, a typical approach is to equip it with language identification (LID) capabilities\cite{watanabe2017language,punjabi2020streaming,zhang2022streaming} . For instance, an additional branch can be introduced in the model for language identification,  and the identified language can be further inputted into the subsequent layers of the model in the form of an embedding or one-hot encoding to provide language information.

\begin{figure}[h]
      \vspace{-0.3cm}
  \centering
  \includegraphics[width=0.8\linewidth]{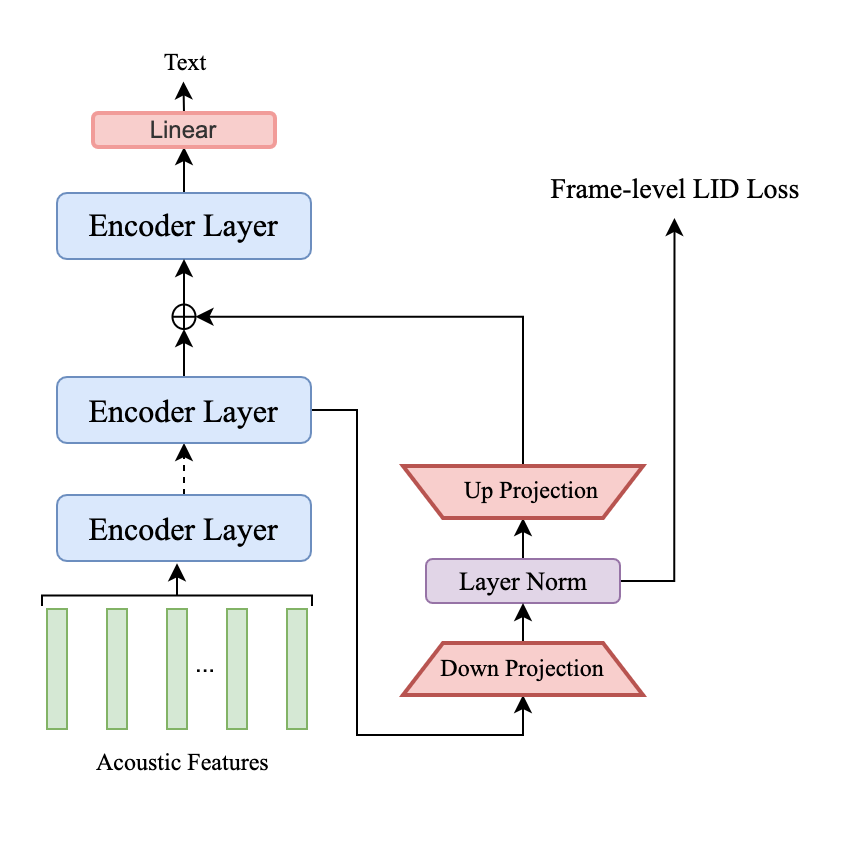}
    \vspace{-0.8cm}
  \caption{The description of frame-level language adapter.}
  \label{fig:speech_production}
\end{figure}

Previous approaches mainly predicted language information at the utterance level through a pooling layer or the last output of LSTM networks. However, relying solely on global language information may not be adequate to equip the model with fine-grained language information. Therefore, we propose the frame-level language adapter (FL-Adapter)  to finely provide language information for each frame of acoustic features. As shown in Figure 2, for parameter efficiency, we designed a structure similar to the adapter\cite{houlsby2019parameter}  network, which introduces a branch from the middle layer of the encoder for frame-level language information prediction. By leveraging a residual connection and incorporating the output of the encoder's intermediate layer,  the language bias information is effectively infused into each frame feature.  For the frame-level LID loss function, cross-entropy loss (expanding LID labels to the same length as the acoustic features) or CTC loss (expanding LID labels to the same length as the text labels) can be used. The total loss function is shown as follows:
\begin{eqnarray*}
&&~~~~~~~~~~~~~~~~~~{L} =  L_{CTC} + {\alpha}L_{LID}~~~~~~~~~~~~~~~~~~~~~~~(4) 
\end{eqnarray*}
where $\alpha$ is a hyperparameter to balance the two losses.

\vspace{-0.1cm}
\subsection{Parameter-efficient Fine-tuning}
\vspace{-0.1cm}
\label{ssec:subhead}
For scenarios that require both language-agnostic and configurable capabilities, we can deploy both models introduced in sections 2.1 and 2.2 simultaneously and choose which model to use based on whether language information is provided in advance. However, this approach can be costly. To address this issue, we attempted to combine the two models using parameter-efficient fine-tuning methods. Specifically, we froze the parameters of the model introduced in section 2.2 and used language prompt tuning, as introduced in section 2.1, for fine-tuning. To further enhance the fine-tuning effect, we also introduced the residual adapter\cite{houlsby2019parameter}  after the feedforward module in each transformer encoder layer. Through these methods, we aim to equip the model with both language-agnostic and configurable capabilities.

\section{EXPERIMENTAL SETUP}
\label{sec:pagestyle}
\subsection{Data Sets}
\label{ssec:subhead}
\vspace{-0.2cm}
To ensure the reproducibility of our proposed methods, we conducted experiments on multilingual open-source datasets. For Chinese, we used AISHELL-1\cite{bu2017aishell}, Aidatatang$\_$200zh\footnote{https://openslr.org/62/} , Magicdata\footnote{ https://openslr.org/68/}, and extracted some data from Wenetspeech\cite{zhang2022wenetspeech}. For Korean, we used the Zeroth-Korean\footnote{https://github.com/goodatlas/zeroth}   dataset. Detailed information on other language datasets can be found in Table 1. Our experimental data is unbalanced in terms of language, which aims to verify whether our methods still work in such a complex situation. The acoustic features we used are 80-dimensional energy-based Log Mel-filterbanks (FBank) computed on a window of 25ms with a 10ms shift.

\vspace{-0.15cm}
\subsection{Modeling Unit}
\vspace{-0.2cm}
Chinese, Japanese, and Korean typically employ characters or words as their modeling units.  However, directly using characters or words as modeling units in multilingual speech recognition would result in a large modeling vocabulary, increasing memory usage and decreasing decoding speed. To address this issue, we converted all languages' texts to the byte level and used a byte pair encoder (BPE) to achieve shared modeling units across different languages. Ultimately, we obtained a multilingual modeling vocabulary of size 6000.

\begin{table}
\caption{Data partitions (in hours)}
\centering
\scalebox{1.0}[1.0]{
\begin{tabular}{llll}
                                &          &                          \\ 
\hline
Language                           & Train & Test   \\ 
\hline
Chinese (ZH)               & 6200     & 10                  \\
English (EN) \cite{panayotov2015librispeech}                & 961          & 5.3                  \\
Japanese (JA) \cite{takamichi2021jtubespeech}                  & 930     & 10           \\
Spanish (ES) \cite{ardila2019common}                  & 435     & 27            \\
Tibetan (BO) \cite{zhao2020open}                     & 84.3     & 3.0            \\
Korean (KO)            & 51.6     & 1.2          \\
Uyghur (UG) \cite{rozi2015open}                   & 21     & 2.7             \\
\hline
Total                   & 8723      & 59.2 \\
\hline
                                &          &         &               
\end{tabular}}
\vspace{-0.6cm}
\end{table}

\vspace{-0.15cm}
\subsection{Implementation Details}
\label{ssec:subhead}
\vspace{-0.2cm}
Our basic  architecture for all experiments is Transformer-CTC, which includes 12 encoder layers with $d^{model}$=768, $d^{ffn}$=3072, and $n^{head}$=12. A convolutional front-end was used to subsample the acoustic features by a factor of 6.

For the baseline system, we trained a language-agnostic model by pooling data from all languages together. For prompt tuning, the prompt encoder is only one embedding layer. For prefix tuning, the prompt encoder consists of one embedding layer and one linear layer, which finally maps to $num$\_$layer$ * $d^{model}$ * $2$ dimensions, and split and concatenated onto each layer's key and value as the prompts.  For the $Concat$ method, we compared using one-hot encoding and embedding, while for the $Add$ and $Attention$ methods, we default to using embedding. For FL-Adapter, the projection dimension is the number of language classes plus one (an extra CTC blank). For the parameter-efficient fine-tuning mentioned in Section 2.3, we may need multiple prompt tokens, this can be achieved by expanding the output layer dimension of prompt tuning and prefix tuning by $num$\_$prompt$ and then splitting it into $num$\_$prompt$ prompt tokens.

All experiments are conducted with the PyTorch toolkit. During training, we adopt the Adam optimizer with  ${\beta _{1}} $= 0.9, ${\beta _{2}} $ = 0.98, and a Noam  learning rate decay strategy with $d$ = 768, $warmup\_step$ = 25000, and $factor$ = 1.0.  \vspace{0.1cm}

\begin{table*}[th]
	\caption{Comparison of  word error rates (WERs, $\%$) for different multilingual speech recognition models.}
\centering
	\scalebox{1.0}[1.0]{%
		\begin{tabular}{lccccccccc}
			\hline
			 \textbf{Model}  &\textbf{Params}     & ~~\textbf{ZH}         & ~~\textbf{EN}    &~~\textbf{JA} &~~\textbf{ES} &~~\textbf{BO} &~~\textbf{KO} &~~\textbf{UG} &~~\textbf{Avg} \\			  
			\hline
			Baseline       &91.6169M    & ~~4.97   &~~ 26.61    &~~14.33 &~~24.41   &~~15.61   &~~13.25     &~~83.18   &~~26.05         \\
			\hline
			\multicolumn{10}{c}{Language-configurable multilingual model} \\
			\hline
            		Add       &91.6174M    & ~~4.93   &~~11.22    &~~14.67 &~~24.03   &~~14.18   &~~15.27     &~~58.22   &~~20.36             \\
		         Attention   &91.6176M   & ~~4.87   & ~~12.89    &~~14.95 &~~25.62   &~~14.71   &~~15.24     &~~59.79   & ~~21.15             \\
		         Concat (one-hot)     &91.6241M   & ~~4.97   &~~10.83    &~~14.85 &~~24.66   &~~15.62   &~~13.97    &~~56.39   &~~20.19             \\
            		Concat (embedding)     &91.6303M   & ~~4.93   &~~10.73    &~~14.86 &~~24.46   &~~15.61   &~~13.98     &~~55.36   &~~19.99             \\
			Prefix Tuning    &91.7927M   & ~~4.68  & ~~11.10     &~~14.58  &~~24.49   &~~13.84    &~~14.16     &~~56.72   &~~19.94             \\
			Prompt Tuning (prefix)   &91.6174M   & ~~4.66  & ~~9.78    &~~14.32 &~~23.76   &~~13.06    &~~11.96     &~~56.95   &~~19.21            \\
			Prompt Tuning (suffix)   &91.6174M   &~~\textbf{4.54}  &~~\textbf{9.33}    &~~14.01  &~~\textbf{22.51}   &~~\textbf{12.42}    &~~ \textbf{11.66}     &~~\textbf{54.17}   & ~~\textbf{18.38}             \\
			Prompt Tuning (both)   &91.6174M   & ~~4.63  &~~10.67     & ~~\textbf{13.96}  &~~23.52   &~~13.84    &~~12.72     &~~56.35   &~~19.39             \\
			\hline
			\multicolumn{10}{c}{Language-agnostic multilingual model} \\
			\hline
			FL-Adapter (CE, $\alpha$=0.2)   &91.6272M   &~~4.85 &~~14.57  &~~14.34  &~~23.88   &~~15.01    &~~13.20     &~~68.19   &~~22.01            \\
			FL-Adapter (CTC, $\alpha$=0.2)   &91.6300M   &~~4.71 &~~12.09  &~~14.30  &~~23.48   &~~14.35    &~~13.14     &~~64.18   &~~20.89            \\
			FL-Adapter (CTC, $\alpha$=0.5)   &91.6300M   &~~\textbf{4.60} &~~\textbf{10.62}  &~~14.14  &~~23.51   &~~14.91    &~~\textbf{12.99}     &~~\textbf{56.70}   &~~\textbf{19.64}            \\
			FL-Adapter (CTC, $\alpha$=0.5)   &98.7171M   &~~4.87 &~~11.02  &~~\textbf{13.76}  &~~\textbf{23.31}   &~~\textbf{14.24}    &~~14.69     &~~63.22  &~~20.73            \\
			FL-Adapter (CTC, $\alpha$=0.8)   &91.6300M   &~~4.85 &~~11.63 &~~14.13 &~~24.41   &~~14.31   &~~15.03    &~~65.28   &~~21.38            \\
			FL-Adapter (CTC, $\alpha$=1.0)   &91.6300M   &~~5.22 &~~13.44  &~~15.26  &~~25.32   &~~15.43    &~~15.26     &~~68.95   &~~22.70          \\
			  \hline
			  \multicolumn{10}{c}{Unified language-agnostic and configurable model} \\
			  \hline
			  FL-Adapter ($\alpha$=0.5, frozen)   &91.6300M &~~4.60 &~~10.62  &~~14.14  &~~23.51   &~~14.91    &~~12.99     &~~56.70   &~~19.64             \\
			  + Prompt Tuning (1 token)   &91.6305M &~~4.60 &~~10.60  &~~14.13  &~~23.50   &~~14.81    &~~12.96     &~~56.70   &~~19.61             \\
			  + Prompt Tuning (5 tokens)   &91.6328M &~~4.66 &~~10.62  &~~14.18  &~~23.48   &~~14.46    &~~12.96     &~~56.60   &~~19.57             \\
			  + Prompt Tuning (10 tokens)   &91.6356M &~~4.60 &~~10.60  &~~14.12  &~~23.48   &~~14.46    &~~12.96     &~~56.55   &~~19.54             \\
			  + Prefix Tuning (1 token)   &91.8328M &~~4.60 &~~10.18  &~~13.87  &~~23.00   &~~13.90    &~~12.54     &~~56.31   &~~19.20             \\
			  + Prefix Tuning (5 tokens)   &92.6438M &~~4.58 &~~9.99  &~~13.79  &~~22.70   &~~13.68    &~~12.11     &~~55.32   &~~18.88           \\
			  + Prefix Tuning (10 tokens)   &93.6576M &~~4.56 &~~9.94  &~~13.77  &~~22.61   &~~13.68    &~~11.99     &~~55.06   &~~18.80             \\
			  + Prefix Tuning (15 tokens)   &94.6713M &~~4.54 &~~9.84   &~~13.76  &~~22.59  &~~13.54    &~~11.99     &~~55.05   &~~18.76             \\
			   ~~~+ Residual Adapter (dim=32)   &95.2338M &~~\textbf{4.52} &~~\textbf{9.59}  &~~\textbf{13.58}  &~~\textbf{22.18}   &~~\textbf{12.91}    &~~\textbf{11.71}     &~~\textbf{54.09}   &~~\textbf{18.37}             \\
			  \hline
	\end{tabular}}
	\vspace{-0.03cm}
\end{table*}

\vspace{-0.2cm}
\section{Results}

\subsection{The Effects of Language Prompt Tuning}
\label{ssec:subhead}
\vspace{-0.2cm}
We compared various methods that utilize prior language information to achieve language-configurable multilingual speech recognition. As shown in Table 2, the prompt tuning-based methods outperformed other methods, while methods such as $Add$ and $Concat$ performed worse than the baseline on some test sets. This can be attributed to the adaptive nature of the prompt tuning-based methods, which enables the model to allocate language information to each frame using self-attention at each layer, as opposed to a blanket imposition on every frame. Additionally, prompt tuning outperforms prefix tuning, because the latter's method of inserting prompts at every layer could potentially result in overfitting. Furthermore, for prompt tuning,  appending the language prompt as a suffix yields superior performance, as it ensures  the continuity of acoustic features and the integrity of semantic content. 

\vspace{-0.1cm}
\subsection{The Effects of Frame-level Language Adapter}
\vspace{-0.2cm}
\label{ssec:subhead}
Table 2  presents a comparison of the two frame-level LID loss, CE (with $\alpha$=0.2 being optimal) and CTC (with $\alpha$=0.5 being optimal) for FL-Adapter. The results  indicate that the CTC loss function yields better performance. This is because speech signals may contain non-linguistic or silent segments, which can be aligned to the blank label using the CTC loss function instead of forcing alignment to the language label. In addition, Table 2 illustrates that FL-Adapter can effectively provide  language bias information, thereby enhancing the performance of multilingual speech recognition.

\vspace{-0.2cm}
\subsection{The Effects of Parameter-efficient Fine-tuning}
\label{ssec:subhead}
\vspace{-0.2cm}
In order to unify language-agnostic and language-configurable scenarios, we froze the optimal FL-Adapter model and fine-tune it using prompt tuning based methods. To demonstrate that our methods are not solely due to an increase in the number of model parameters, we also trained a FL-Adapter model with increased parameters (13 layers) as a new reference. As shown in Table 2,  prompt tuning does not lead to significant improvement, as it only adds learnable language prompt tokens to the input layer with limited fine-tuning ability. In contrast, prefix tuning introduces language prompt tokens in every Transformer layer, resulting in greater fine-tuning capability, and the performance of prefix tuning improves as the number of prompt tokens increases. In addition, we added residual adapter networks to each layer of the Transformer encoder upon prefix tuning, which further improves the speech recognition performance. This is owing to prefix tuning mainly fine-tunes the self-attention module of  Transformer layers, while the residual adapter network complements and fine-tunes the feedforward module of  Transformer layers, and the  integration of the two culminates in superior outcomes. Finally, we improved the performance of the FL-Adapter  model to be comparable to the language-configurable model through these  parameter-efficient fine-tuning methods, this allows the model to perform well across different languages while still being configurable for specific languages.

\section{Conclusions}
\label{ssec:subhead}
\vspace{-0.2cm}
This paper proposed two parameter-efficient methods to enhance the performance of CTC-based end-to-end multilingual speech recognition in both language-agnostic and language-configurable scenarios. We also explored some parameter-efficient fine-tuning methods to unify the two scenarios into a unified model. The proposed methods achieved significant performance improvements across seven languages with imbalanced data. In the future, we plan to further validate our methods on larger models and a wider range of languages.

\vfill\pagebreak
\bibliographystyle{IEEEbib}
\bibliography{strings,refs}

\end{document}